\documentclass{aastex}
\usepackage{spr-astr-addons}
\usepackage{url}
\usepackage{subfigure}
\RequirePackage{color}
\def\image1{\centerline{\color[gray]{.75}\rule{\hsize}{4pc}}}%

\begin{document}

\title{Interacting New Agegraphic Dark Energy in a Cyclic Universe}
\slugcomment{Not to appear in Nonlearned J., 45.}
\shorttitle{Short article title}

\shortauthors{K. Saaidi et al.}

\author{K. Saaidi$^{*}$\altaffilmark{1}}  \and \author{H. Sheikhahmadi$^{*}$ \altaffilmark{2}} \and \author{A. H. Mohammadi$^{**}$ \altaffilmark{3}}
\affil{$^{*}$Department of Physics, Faculty of Science, University of
Kurdistan,  Sanandaj, Iran.} \and \affil{$^{**}$Islamic Azad University, Evaz Branch, Fars Province, Iran}


\altaffiltext{1}{ksaaidi@uok.ac.ir}
\altaffiltext{2}{h.sh.ahmadi@uok.ac.ir}
\altaffiltext{3}{abolhassanm@gmail.com}

\begin{abstract}
The main goal of this work is investigation of NADE in the cyclic universe scenario.
Since, cyclic universe is explained by a phantom phase
($\omega<-1$), it is shown when there is no interaction between
matter and dark energy, ADE and NADE do not produce a phantom
phase, then can not describe cyclic universe. Therefore, we study
interacting models of ADE and NADE in the modified Friedmann equation. We find out that, in the high
energy regime, which it is a necessary part  of cyclic universe
evolution, only NADE can describe this phantom phase era for cyclic
universe. Considering deceleration parameter tells us that the universe has a deceleration phase after
an acceleration phase, and NADE is able to produce a cyclic universe. Also it is found valuable to study
generalized second law of thermodynamics. Since the
loop quantum correction is taken account in high energy regime, it may not be suitable to use standard
treatment of thermodynamics, so we turn our attention to the result of \citep{29}, which the authors
have studied thermodynamics in loop quantum gravity, and we show that which condition can satisfy
generalized second law of thermodynamics.
\end{abstract}




\section{Introduction}

Cosmological and astronomical observations such as supernovae type
Ia observational data \citep{1, 1aa, 1ab} and Wilkonson Microwave
Anisotropic Probe (WMAP) \citep{2, 2aa, 2ab, 2ac, 2ad}  imply that the universe is
undergoing a period of accelerated expansion. Since normal matter can not give rise to accelerated
expansion of the universe, Scientists came up with a solution which expresses that this expansion is a
result of an ambiguous fluid named Dark Energy  \citep{3, 3aa}. \\
The theoretical and experimental analysis suggest that the universe
consist of $\%73$ dark energy, $\%23$ cold dark matter (CDM), and
remnant matter is baryons \citep{4, 4aa}. Unfortunately the nature and
origin of dark energy are ambiguous up to now, but people have
proposed some candidates to describe dark energy. Amongst the
various candidates of dark energy to describe accelerated expansion
of the universe, cosmological constant (vacuum energy), $\Lambda$,
with equation of state (EoS) $\omega=-1$ is located in central
position. However, as it is well known, the cosmological constant
proposal has two famous problems, fine-tuning problem  and the cosmic
coincidence problem \citep{5, 5aa, 5ab, 5ac}. Some other of dark energy models suggest
that dark energy component can treat as scalar field with dynamical
EoS. In this scenario the evolution of the field is very slow, so
that kinetic energy density is less than the potential energy
density, and this give us a negative pressure, responsible to the
cosmic acceleration \citep{6, 6aa, 6ab}. Some of scalar field models are as
chameleon field \citep{7, 7aa}, quintessence (Q-field) \citep{8, 8aa}, and phantom. In the phantom
field scenario, the parameter of EoS is as $\omega<-1$, due to
existence of a negative kinetic energy density of scalar field. It
is well known that the phantom dark energy model suffers from two
kind of problems, "Big Bang" singularity and "Big Rip" singularity,
where big bang is related to initial epoch of universe and big rip
is related to a finite future singularity. Since the space-time singularities are invidious
for theorists, models that avoid these singularities are attractive. One of these models
names cyclic universe which have received huge attention \citep{15, 15aa, 15ab}. Presence of
$\rho^2$ term with a negative sign in Friedmann equation (which is used at studying cyclic universe)
is an effective way to eliminate these singularities.  In the cyclic universe
scenario, where is based on the phantom dark energy model, the
universe oscillates through a series of expansion and contraction.
Universe in this scenario has a very high energy density at
beginning an ending of the expansion, so quantum gravity can not be
ignored in these stages \citep{16, 16aa}. This evolution can be result
from the modified Friedmann equation in the loop quantum cosmology
(LQC). In LQC, the Friedmann equation has been modified to
\begin{equation}
H^2=\frac{\rho}{3m_p^2}\left( 1-\frac{\rho}{\rho_c} \right),
\end{equation}
where $H$ is the Hubble parameter, $m^2_{p}$ is the reduced planck
mass ($m_p^2=\frac{1}{8\pi G}=2.44\times 10^{18}GeV$), $\rho$ is
the total of energy density, $\rho_c$ is the critical energy
density as $\rho_c=4\sqrt{3}\gamma^{-3}m_p^4=0.82 \rho_{p}$,
where $\rho_{p}=2.22\times 10^{76}GeV$, and $\gamma$ is the
dimensionless Barbero-Immirizi parameter \citep{17}. We notice,
this correction can solve the singularity problems as follow,
when the total energy density reaches the critical density, the
universe reaches the maximum at the end of expansion that is called "turnaround point",
and  universe arrives at smallest size at the end of contraction, then we have a bounce there
\citep{18, 18aa, 18ab, 18ac}. We emphasize the idea of cyclic universe was first
introduced by Tolman \citep{19, 19aa}.\\
An interesting attempt for probing the nature of dark energy, in the
framework of quantum gravity, is the holographic dark energy (HDE).
In the HDE model, dark energy is a dynamical evolving vacuum energy
density that can satisfy the phantom behavior. Authors of
\citep{20} have investigated the cyclic universe by HDE (and some
interesting work about HDE have been done \citep{21, 21aa, 21ab,21ac, 22, 22aa, 23}.
Another attractive model to describe the nature of dark energy,
within the framework of a fundamental theory originating from some
considerations of the feature of quantum gravity theory, is called
agegraphic dark energy (ADE) model \citep{24}. The ADE assumes that
the dark energy comes from the universe components fluctuation such
as space-time and matter fluctuation (for further discussion we
refer the reader to \citep{25, 25aa, 26, 26aa, 26ab}. In this model, the age of
universe is taken as the length measure instead of the horizon
distance, therefore the causality problem which appears in the HDE
model can be avoided. The ADE model suffers from the difficulty to
describe the matter dominant epoch. The authors of \citep{27} have
introduced a new mechanism to overcome that problem, which it is
called new agegraphic dark energy (NADE) model, and its energy
density is defined by $\rho_{\Lambda}=3n^2m^2_p \eta^{-2}$, where
$3n^2$ is introduced to parameterized some uncertainties  and $\eta$
is conformal time and can be written as
\begin{equation}
\eta=\int_0^a \frac{da}{a^2H},
\end{equation}
where $a$ is scale factor and
$H$ is well-known as Hubble parameter. Cyclic model of universe,
due to avoiding singularity, and NADE, due to estimating
a good approximate of dark energy value and solving causality
problem of HDE, have received huge interest.
 In previous works, like \citep{28-a}, it was only mentioned that interaction NADE can produce phantom, however it was not explained that if NADE can produce cyclic universe and how the
quantities like dark energy density parameter, equation of state parameter, and deceleration parameter behave in cyclic universe by taking NADE as component of dark energy.
 In this work we motivated to take NADE as dark energy and investigate universe evolution in cyclic model. If it is able to stand in phantom area, it may produce
cyclic universe, however it is not all of story. Although, presence of this type of dark energy
displays an accelerated universe, but the most important thing in cyclic universe is that, after a while, the
universe should enter in a deceleration area and recontract in turnaround point.\\

\section{High and low energy}

In this step, we focus on high energy regime. Certainly the quadratic term of energy density
in the Friedmann equation can not be ignored, this term can play
a very impressive role in the evolution of universe. The modified
Friedmann equation is as

\begin{equation}\label{9}
3H^2=\rho \left( 1-\frac{\rho}{\rho_c} \right),
\end{equation}
where $\rho$ is a combination of dark energy density, $\rho_{\Lambda}$, and matter
density, $\rho_m$. For dark energy dominant in high energy regime, the above relation
can be rewritten as

\begin{equation}\label{10}
3H^2 \approx \rho_{\Lambda} \left( 1-\frac{\rho_{\Lambda}}{\rho_c}
\right),
\end{equation}
so, according to this relation the parameter of dark energy may
be estimated as

\begin{equation}\label{11}
\Omega_{\Lambda}^{\natural}=\frac{\rho_{\Lambda}}{3H^2} \approx
\frac{\rho_c}{\rho_c-\rho_{\Lambda}}.
\end{equation}
$\Omega_{\Lambda}^{\natural}$ is always larger than one, and it
can be very large when the universe approaches to the turnaround
point. However note that, since we have $\rho_m$, although it is very
small, $\Omega_{\Lambda}^{\natural}$ can not be infinite. This small
value of $\rho_m$ does not allow
$\rho_{\Lambda}$ to reach the exact value of $\rho_c$, then the
presence of $\rho_m$ can prevent the infinite value of
$\Omega_{\Lambda}^{\natural}$.\\
If we assume there is no interaction between these  two
components of the universe, conservation equation can be written as

\begin{eqnarray}\label{1-a}
\dot{ \rho}_m +3H(1+\omega_m)\rho_{m}&=&0 \nonumber \\
\dot{\rho}_{\Lambda} +3H(1+\omega_{\Lambda})\rho_{\Lambda}&=&0.
\end{eqnarray}
By taking ADE, which is defined as $\rho_{\Lambda}=3n^2 T^{-2}$ where $T$ denotes time,
as dark energy component of universe, $\omega_{\Lambda}$ is obtained as

\begin{equation}\label{12}
\omega_{\Lambda}=-1+\frac{2}{3n}\sqrt{\Omega_{\Lambda}^{\natural}}.
\end{equation}

Also, if we replace NADE instead of ADE, we arrive at same result but only scale
factor, $a$, should be added in the denominator of second term on the right hand of
relation (\ref{12}). Since the latest term, in both case, is always positive
$\omega_{\Lambda}$ never can be smaller than $-1$. Since phantom type
of dark energy is requisite for cyclic universe (phantom energy density is getting
larger with increasing scale factor so the total energy density reaches the critical
energy density in turnaround point and universe begins contraction), selecting
ADE and NADE, without interaction, as a component of dark energy in cyclic
universe is not suitable \citep{28-a}.
\par Assumption of interaction between $\rho_{m}$ and $\rho_{\Lambda}$ may
solve above problem. By including interaction, the conservation equations are as

\begin{eqnarray}\label {14}
\dot{ \rho}_m +3H(1+\omega_m)\rho_{m}&=&Q\\
\dot{\rho}_{\Lambda} +3H(1+\omega_{\Lambda})\rho_{\Lambda} &= & -Q,
\end{eqnarray}
where $Q$ indicates interaction. $Q$ is taken as
$Q=\Gamma\rho_{\Lambda}$, with $\Gamma >0$, which means there is
transfer of energy from $\rho_{\Lambda}$ to $\rho_m$ {\cite{28}}.
We take $Q$ as $Q=3b^2H(1+r)\rho_{\Lambda}$, where
$r=\frac{\rho_m}{\rho_{\Lambda}}$. Setting ADE as the component of dark energy gives us an $\omega_{\Lambda}$ as

\begin{equation}\label {15}
\omega_{\Lambda}=-1+\frac{2}{3n}\sqrt{\Omega_{\Lambda}^{\natural}}-b^2(1+r).
\end{equation}
$\Omega_{\Lambda}^{\natural}$ is always larger than one, and it
becomes very large at turnaround point, so for having
$\omega_{\Lambda}<-1$, we should have
$b^2(1+r)>\frac{2\sqrt{\Omega_{\Lambda}^{\natural}}}{3n}$. This
predicts a large value for coupling constant $b$, while it is in contrast to the obtained
value of another papers such as (\citep{28}). \\
Now, NADE is taken as $\rho_{\Lambda}$ which that is the main case of this
work. Because we use conformal time $\eta$ instead of $T$, scale factor appear in relation (\ref{15}), namely

\begin{equation}\label {16}
\omega_{\Lambda}=-1+\frac{2}{3na}\sqrt{\Omega_{\Lambda}^{\natural}}-b^2(1+r).
\end{equation}
To have $\omega_{\Lambda}<-1$, the coupling constant $b$ should
obey following relation

\begin{equation}\label {17}
b^2>\frac{2}{3na}\sqrt{\Omega_{\Lambda}^{\natural}},
\end{equation}
here, $r$ has been ignored because of $\rho_m\ll\rho_{\Lambda}$.
If $\Omega_{\Lambda}^{\natural}$ be in order of $a^2$, we may
obtain a convenient value for $b$ in order it could give us
phantom dark energy in this regime.

\begin{figure}[ht]
\centerline{\includegraphics[width=8cm]{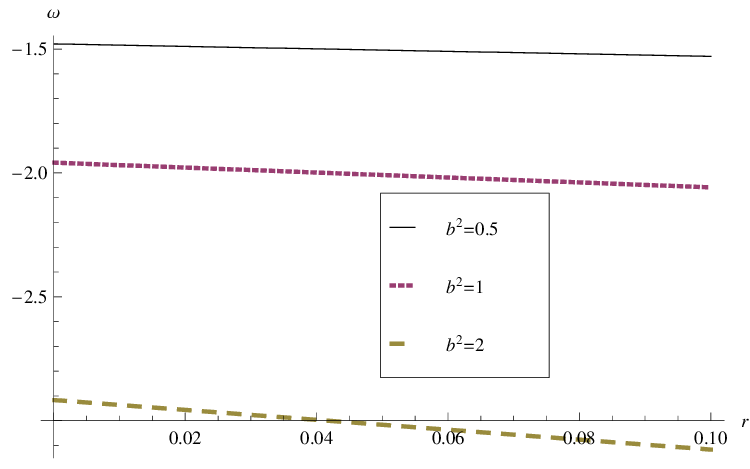}}
\caption{\label{Fig1}\small equation of state parameter of dark energy has been plotted versus the dimensionless quantity $r=\frac{\rho_m}{\rho_{\Lambda}}$, for three different value of interaction coupling constant. }
\end{figure}

 In Fig.1 $\omega_{\Lambda}$ parameter has been plotted versus $r=\frac{\rho_m}{\rho_{\Lambda}}$, for three different value of interaction coupling constant. In expansion phase, by passing time, the quantity $r$ decreases and makes $\omega_{\Lambda}$ larger. Since it still stands in phantom range, larger value of $\omega_{\Lambda}$ shows that dark energy density grows up slower; in contrast to the contraction phase. Also, having stronger interaction coupling causes smaller value for $\omega_{\Lambda}$.

If $\Omega_{\Lambda}^{\natural}$ be in order of $a^{\xi}$, where
$\xi \leq 2$, equation (\ref{17}) can be valid in good approximation.
Now suppose $\Omega_{\Lambda}^{\natural}$ is in order of $a^2$, for
instance $\Omega_{\Lambda}^{\natural}=\alpha a^2$. From equation
(\ref{17}) we have
$b^2>\frac{2\sqrt{\alpha}}{3n}$, and by taking the value of \citep{28} for $b$, namely
$b^2=0.25$, we obtain the value of $\alpha$ as
$\sqrt{\alpha}<\frac{3}{8}n$.\\
Now, we want to obtain differential equation for
$\Omega_{\Lambda}^{\natural}$. In the NADE, $\rho_{\Lambda}$ is
given as

\begin{equation}\label {17-a}
\rho_{\Lambda}=\frac{3n^2}{\eta^2}.
\end{equation}
From the definition of dark energy density parameter, namely
$\Omega_{\Lambda}^{\natural}=\frac{\rho_{\Lambda}}{3H^2}$, The
differential equation for $\Omega_{\Lambda}^{\natural}$ is
acquired as

\begin{equation}\label {17-b}
{\Omega_{\Lambda}^{\natural}
}'=-2\Omega_{\Lambda}^{\natural}\left(
\frac{\dot{H}}{H^2}+\frac{\sqrt{\Omega_{\Lambda}^{\natural}}}{an},
\right)
\end{equation}
where prime denote derivative with respect to ${\cal N}= \ln a$.
Taking the time derivative of modified Friedmann equation, and
substituting that in the above equation, we obtain

\begin{equation}\label {17-c}
\frac{\dot{H}}{H^2}=\frac{-3}{2}\left(
(1+\omega_m)\Omega_m^{\natural} +
(1+\omega_{\Lambda})\Omega_{\Lambda}^{\natural} \right)\left(
1-\frac{2\rho}{\rho_c} \right),
\end{equation}
since, in the high energy regime, $\Omega_m^{\natural}$ can be
ignored against to $\Omega_{\Lambda}^{\natural}$, therefore one
can estimate

\begin{equation}\label {17-d}
\frac{\dot{H}}{H^2} \approx \frac{-3}{2}
(1+\omega_{\Lambda})\Omega_{\Lambda}^{\natural} \left(
1-\frac{2\rho}{\rho_c} \right),
\end{equation}
(for driving above equation we have used
$\frac{H^2}{\rho_c}=\frac{\Omega_{\Lambda}^{\natural}-1}{3{\Omega_{\Lambda}^{\natural}}^2}$,
see ref.\citep{28}. The differential equation, which governs the
NADE evolution of universe in high energy regime, can be attained as

\begin{equation}\label {17-e}
{\Omega_{\Lambda}^{\natural} }'=-2\Omega_{\Lambda}^{\natural}
\Bigg\{
\frac{\sqrt{\Omega_{\Lambda}^{\natural}}}{na}(\Omega_{\Lambda}^{\natural}
-1 )-3b^2(1+r)(\frac{1}{2}\Omega_{\Lambda}^{\natural}-1) \Bigg\}
\end{equation}
with attention to (\ref{17}), it is clearly shown that ${\Omega_{\Lambda}^{\natural}}'=\frac{1}{H}\dot{\Omega}_{\Lambda}^{\natural}$ is positive. So one can
realize that when the universe is growing up (decreasing) $\dot{\Omega}_{\Lambda}^{\natural}$ is positive (negative) which
it displays that dark energy density parameter is increasing (decreasing) by passing time. By passing the universe from
turnaround point, $\Omega_{\Lambda}^{\natural}$ decreases and puts the universe in low energy regime.
\par Another useful cosmological parameter is deceleration
parameter which can tell us if the expansion of universe is accelerating or not. Deceleration parameter is given by

\begin{equation}
q=-1-\frac{\dot{H}}{H^2},
\end{equation}
substituting $\frac{\dot{H}}{H^2}$ term in the above relation, we
arrive at

\begin{eqnarray}
q&=&-1+\frac{3}{2}\left(
(1+\omega_m)\Omega_m^{\natural}+(1+\omega_{\Lambda})
\Omega_{\Lambda}^{\natural}
 \right) \nonumber \\
 & & \ \ \ \ \ \ \ \times \left(1-\frac{2\rho}{\rho_c}
  \right) \nonumber \\
  & =&-1+\frac{3}{2}(1+\omega_{\Lambda})\Omega_{\Lambda}^{\natural}
  \left(1-\frac{2\rho}{\rho_c}
  \right).
\end{eqnarray}
\begin{figure}[ht]
\centerline{\includegraphics[width=8cm]{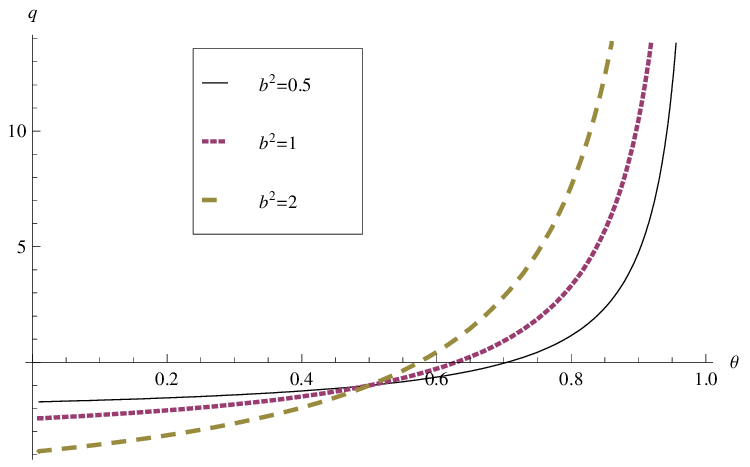}}
\caption{\label{Fig2}\small (expansion epoch) Deceleration parameter, $q$, has been plotted versus the dimensionless quantity $\theta=\frac{\rho_{\Lambda}}{\rho_c}$, for three different value of interaction coupling constant. }
\end{figure}
\begin{figure}[ht]
\centerline{\includegraphics[width=8cm]{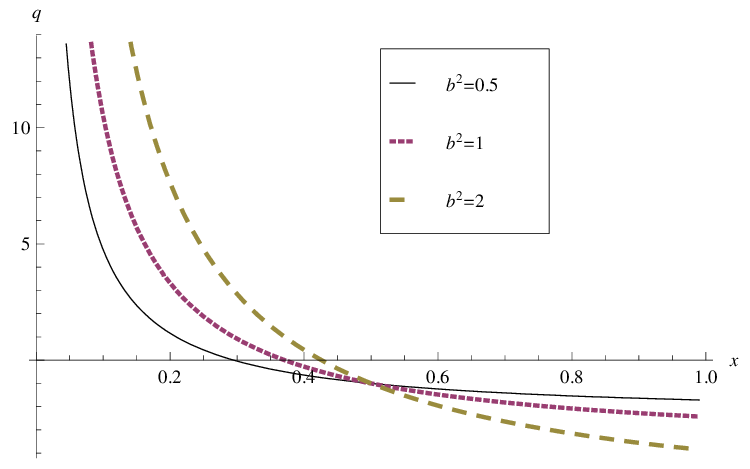}}
\caption{\label{Fig3}\small (contraction epoch) Deceleration parameter, $q$, has been plotted versus the dimensionless quantity $x=1-\theta$, for three different value of interaction coupling constant. }
\end{figure}
 Deceleration parameter has been plotted versus the dimensionless quantity $\theta = \frac{\rho_{\Lambda}}{\rho_c}$ for expansion and contraction epoch of universe evolution in Fig.2. and Fig.3. respectively (with attention to this fact that $\omega_{\Lambda}$ stands in phantom range). In the plots three different values for $b$ are taken as $b^2=0.5, 1, 2$; which displays the strange of interaction between matter and dark energy. Fig.2. shows that $q$ has negative values that it means the universe is in an acceleration phase and getting larger. Since there is a positive value for Hubble parameter, $\Omega_{\Lambda}^{\natural}$ increases too. By passing time and increasing dark energy density, $q$ approaches to zero, therefore the universe moves from acceleration phase to deceleration phase. However since we are in expansion epoch of universe evolution, $\Omega_{\Lambda}^{\natural}$ is still getting larger which causes $q$ grows rapidly (and as we expect larger values of interaction coupling constant makes $q$ to grow up faster). Reaching energy density to critical energy density, the universe is forced to stop expansion and change its evolution direction and start to contract in turnaround point. In contraction epoch, the universe is getting smaller more and more with a large value of $q$ at the beginning of contraction, as it has been shown in Fig.3. The Hubble parameter has negative values and $\Omega_{\Lambda}^{\natural}$ decreases therefore dark energy density is diluted. $q$ again takes negative values, however the universe continue to contract and decrease more and more, until it moves to low energy regime and dark energy dilute enough in which matter and radiation become dominant.


\par Up to now, from above discussion in high energy regime, we found
out that NADE can satisfy a
phantom fluid for cyclic scenario. However this is not at all our universe,
there is also regime related to
the low energy, where $\rho \ll \rho_{c}$. At this regime we have the
usual Friedmann equation for
universe, namely $$3H^{2}=\rho=\rho_{m}+\rho_{\Lambda}.$$ By
defining the energy density parameters in the form of

\begin{equation}
\Omega_{i}=\frac{\rho_{i}}{3H^{2}},
\end{equation}
where $i$ refer to matter and dark energy. One can rearrange the
standard Friedmann equation as $1=\Omega_{m}+\Omega_{\Lambda}$,
from this relation we find out the maximum value of $\Omega_{i}$
is one. Now, NADE is taken as dark energy component with interaction
with matter, so from
conservation equation related to the dark energy and with the help of above Friedmann equation, one
obtain $\omega_{\Lambda}$ as

\begin{equation}
\omega_{\Lambda}=-1+\frac{2}{3a}\frac{\sqrt{\Omega_{\Lambda}}}{n}-3b^{2}(1+r),
\end{equation}
maximum value of dark energy density parameter, namely $\Omega_{\Lambda}$ is one and also
when the scale factor is large enough to make the second term on the right hand of above relation
small in which
$\omega_{\Lambda}$ stands in phantom range. The sort of dark energy as a component of universe
fluid has been determined, so we want to be aware about the evolution of universe in this stage. \\
The deceleration parameter is expressed as

\begin{equation}
q=-1+\frac{3}{2} \Big( (1+\omega_{m})\Omega_{m}+(1+\omega_{\Lambda})\Omega_{\Lambda} \Big),
\end{equation}
so in the earlier time when $\Omega_{\Lambda}$ converge to zero (and $\Omega_{m}$ converge to
one) the deceleration parameter is as $q\approx 1$ for $\omega_{m}=\frac{1}{3}$, and $q\approx
\frac{1}{2}$ for $\omega_{m}=0$ that indicates a deceleration universe. Whereas in the late time
when $\Omega_{\Lambda}$ approaches to one , the deceleration parameter has negative sign that
introduces an acceleration universe. \\
\section{Validity of second law of thermodynamics}

So far, we investigated whether NADE can produce a phantom dark
energy for cyclic universe. Now, it can be valuable to study validity of
generalized second law of thermodynamic in the model and consider which condition can satisfy
generalized second law. Since loop quantum
gravity correction enters in Friedmann equation which plays an important role in
the evolution of universe, especially in high energy regime, it is not suitable one
straightforward extends the standard treatment of thermodynamic. For this
reason we use the result of \citep{29} which thermodynamic has been
investigated in loop quantum cosmology there. The authors of \citep{29}
have performed their work by taking a dynamical apparent horizon as universe
horizon. It has been demonstrated that one can have
\begin{equation}\label {20}
dE=TdS+WdV.
\end{equation}
However note that, here $E=\rho_{eff}V=\rho \left( 1-\frac{\rho}{\rho_{c}}
 \right)V$ which $\rho_{eff}$ is effective energy density and $W=\frac{1}{2}(\rho_{eff}-P_{eff})$ (in which $P_{eff}=P \left( 1-\frac{2\rho}{\rho_{c}}
 \right)  - \frac{\rho^2}{\rho_c^2}$ is effective pressure). The relation can be very useful in this step
 of the work. Like \citep{29}, we take apparent horizon $$R_{A}=\frac{1}{H}.$$
 From considering the validity of generalized second law, we should sum variation of
 entropy for both of horizon and fluid inside the horizon. The horizon entropy

\begin{equation}
\dot{S}_h=2\pi R_{A}\dot{R}_{A}
\end{equation}
For the relation (\ref{20}), the entropy variation of fluid inside the horizon can be acquired as

\begin{equation}
T_{I}\dot{S}_{I}=V\dot{\rho}_{eff}+\frac{(\rho_{eff}+P_{eff})}{2} \dot{V}
\end{equation}
Note that, there is an equilibrium between the temperature of horizon and matter
inside the horizon. Using modified Friedmann equation and conservation relations
and also with the help of the definition of horizon and energy density parameters,
the total entropy variation is resulted as

\begin{eqnarray}
\dot{S}_{tot}&=&3\pi \Big( (1+\omega_m)\Omega_m +
(1+\omega_{\Lambda})\Omega_{\Lambda}^{\natural} \Big)
\left( 1-\frac{2\rho}{\rho_c} \right) \nonumber \\
 & & \Bigg\{ R_{A} + \frac{3}{T_I} \Big((1+\omega_m)\Omega_m +
 (1+\omega_{\Lambda})\Omega_{\Lambda} \Big) \nonumber \\
  & & \times \left( 1-\frac{2\rho}{\rho_c}  \right) - \frac{4}{T_I} \Bigg\}
\end{eqnarray}
Now, we turn our attention to the high energy regime where $\Omega_m$ can be ignored
against to the larg value of $\Omega_{\Lambda}^{\natural}$. Define
$\alpha=\frac{\rho}{\rho_c}$ in which $0<\alpha<1$, then the total entropy variation is
reorganized as

\begin{eqnarray}
\dot{S}_{tot}&=&3\pi (1+\omega_{\Lambda})(1-2\alpha)
\Omega_{\Lambda}^{\natural} \nonumber \\
 & & \times \Bigg\{ R_{A} + \frac{3}{T_I}(1+\omega_{\Lambda})(1-2\alpha)
\Omega_{\Lambda}^{\natural} - \frac{4}{T_I}\Bigg\},
\end{eqnarray}
where $(1+\omega_{\Lambda})<0$. If $\alpha<\frac{1}{2}$, for making valid the
second law, there should be

$$R_A<-\frac{3(1+\omega_{\Lambda})(1-2\alpha)}{T_I}
\Omega_{\Lambda}^{\natural}.$$
However, if $\alpha>\frac{1}{2}$, the total entropy variation is always positive.\\
In low energy density, $\Omega_{\Lambda}^{\natural}$ tends to
$\Omega_{\Lambda}$ where $0<\Omega_{\Lambda}<1$, also $\alpha$ approaches
to zero, so one arrives at

\begin{equation}
\dot{S}_{tot}=3\pi \beta \left( R_A + \frac{3\beta}{T_I} - \frac{4}{T_I} \right)
\end{equation}
where $\beta=\Big( (1+\omega_m)\Omega_m + (1+\omega_{\Lambda})
\Omega_{\Lambda} \Big)$. In radiation and matter dominant,
where $\Omega_{\Lambda}$ converges to zero, $\beta$ is positive and we should have

$$T_I>\frac{4-3\beta}{R_A}$$
to generalized second law be valid. In radiation and matter dominant $\omega_m$
is equal to $\frac{1}{3}$ and $0$ respectively. This expresses that the right term is
positive.  Passing time, the dark energy eventually dominates and
$\Omega_{\Lambda}$ increases to one, so $\beta<0$. The generalized second law
leads us to the below condition for temperature $$T_I \leq \frac{4-3\beta}{R_A}$$
and since $\beta$ is negative we can have a positive value for temperature. So the generalized
second law of thermodynamics can be valid, and it can be a confirmation for the model.


\section{Conclusion}
During this work, in brief, it was shown that both of ADE and NADE, without interaction
with matter in both high and low energy regime, can not satisfy phantom dark energy which it is a necessary condition for cyclic universe
scenario. So, as it was the main goal of the work, an interaction between the fluid components of the
universe was supposed. At first step, we turned our attention to the high energy regime, and we
realized that , in ADE case for providing a phantom fluid there must be a large value for coupling
constant $b$ which it is not compatible with the results of previous works, but NADE can produce a
phantom fluid for cyclic universe while dark energy density parameter, namely
$\Omega_{\Lambda}^{\natural}$, behaves as $a^{\xi}$ where $\xi \leq 2$.
 It was explained that, in expansion phase, equation of state parameter of dark energy increases by decreasing $r=\frac{\rho_m}{\rho_{\Lambda}}$. However in contraction phase, the quantity $r$ increases, so $\omega_{\Lambda}$ decreases. In contraction phase with attention to this fact that dark energy stands in phantom range, smaller values of $\omega_{\Lambda}$ express that dark energy density decreases faster and let universe to enter matter dominant phase.  Moreover larger value of interaction coupling constant produce smaller valus for $\omega_{\Lambda}$.\\
Considering differential equation of $\Omega_{\Lambda}^{\natural}$ exhibited that after passing the universe
from turnaround point (in contracting phase), $\Omega_{\Lambda}^{\natural}$ decreases and after a while the universe
is putted in low energy regime, then matter and radiation are be allowed to be dominate.
 By computing deceleration parameter, it was found out that there is negative values for $q$ which shows that the universe is in an accelerated expansion phase. Increasing dark energy density enters $q$ in positive area and it is made larger. Then, the universe moves from an accelerated expansion phase to a deceleration phase and it is slowing down until it stops expansion and starts contraction in turnaround point. So it is realized that the model is able to create a deceleration phase and start contraction after an accelerated expansion phase.
In next step, we investigated the
low energy regime of universe where it can be possible to ignore loop quantum correction in
Friedmann relation and go back to usual form. Investigation shew us that, in low energy regime,
$\omega_{\Lambda}$ stands in phantom range. Also, considering deceleration parameter displayed
that universe expansion decelerates in both of matter and radiation dominant, where
$\Omega_{\Lambda}$ is small enough, and accelerates by increasing $\Omega_{\Lambda}$ toward
one. In the last part of the work, we studied the validity of generalized second law of thermodynamic in both
case of high and low energy. In high energy it was determined that when $\alpha=\frac{\rho}{\rho_c}>\frac{1}{2}$
the total entropy variation is always positive. At low energy, we shew that to make valid the law,
a bound for temperature is imposed.


\end{document}